\documentclass[preprint,showpacs,preprintnumbers,amsmath,amssymb]{revtex4}

\usepackage{graphicx}% Include figure files
\usepackage{dcolumn}% Align table columns on decimal point
\usepackage{bm}% bold math

\begin{document}

\title{Early-time interface instabilities in high intensity 
\\aero-breakup of liquid drop}% Force line breaks with \\

\author{X. Y. Hu} \author{N. A. Adams}
\affiliation{%
Lehrstuhl f\"{u}r Aerodynamik und Str\"{o}mungsmechanik, Technische Universit\"{a}t
M\"{u}nchen
\\ 85748 Garching, Germany}%

\date{\today}% It is always \today, today,
             %  but any date may be explicitly specified

\begin{abstract}
The early-time interface instabilities in high intensity (high Weber number and high Reynolds number) 
aero-breakup of a liquid drop are investigated by numerical simulations.
A combined analysis based on simulation results and linear-instability theory show 
that both RT (Rayleigh-Taylor) and KH (Kelvin-Helmholtz) instabilities 
contributes the dominant disturbances originate from 
about the half way from the stagnation point to the equator.
This is verified further with a specially modified simulation, 
which decreases the effect of KH instability while keeps other flow properties unchanged.
\end{abstract}

\maketitle

The deformation and shattering 
of liquid drop after being suddenly exposed to uniform high speed gas flow 
(usually behind a shock wave) is called aerobreakup, 
which has a great variety of practical applications,
such as internal combustion engines, 
nuclear fusion devices, agricultural and environmental sprays.
Aerobreakup is the subsequence of physical instabilities. 
For high intensity (high Weber number and high Reynolds number) 
aerobreakup of a Newtonian liquid drop, 
the weak viscous dissipation and surface tension 
allow rapid growth of high-wavenumber disturbances. 
Due to the extremely high acoustic impedance mismatch at the interface, 
Richtmeyer-Meshkov (RM) instability is negligible.
Therefore, the Kelvin-Helmholtz (KH) instability and the Rayleigh-Taylor (RT) instability 
are proposed as the main mechanisms. 
It was first postulated that the exponentially growing RT instability 
at the upstream front of the drop lead to a catastrophic shattering of the drop 
\cite{harper1972breakup, joseph1999breakup}.
Based on their experimental observations and simplistic estimations,
Theofanous et al. \cite{theofanous2004aerobreakup, theofanous2008physics, theofanous2011aerobreakup}
argued that the dominant mechanism is the shear-induced motion with a significant radial component and instabilities on the so-generated, stretched liquid sheet (more like KH instability).
However, since the spatial/temporal resolutions of the up-to-date video camera 
(about 200 pixel per millimeter/50 kHz \cite{theofanous2011aerobreakup})
is still not able to resolve the high intensity aerobreakup,
it is difficult to verifying these mechanisms by merely experimental approaches.

Here, we investigate these two mechanisms at the early time of high intensity aerobreakup by numerical simulations. 
The mathematical idealization of this problem is that of a two-dimensional, 
compressible Newtonian flow with vanishing surface tension and viscous dissipation. 
The governing equation can be written as a system of conservation laws
\begin{equation}
\frac{\partial \mathbf{U}}{\partial t}+\nabla \cdot \mathbf{F} = 0 \quad {\rm on} ~\Omega,
\label{conservation-law}
\end{equation}
where $\Omega$ is the flow domain, which is separated by the 
phase interface into two sub-domains of gas phase $g$ and liquid phase $l$. 
$\mathbf{U}=(\rho, \rho\mathbf{v}, E)^{T}$ is the density of the conserved quantities, 
such as mass, momentum and energy, $\mathbf{F}$ represents the corresponding flux functions.
The equation of states for the gas and liquid are the ideal-gas equation and Tait's equation, respectively.
Eq. (\ref{conservation-law}) is solved by the conservative sharp-interface method 
\cite{HuKhooAdmasHuang2006, hu2009hrs}.
In this method, a standard finite volume approach using the 5th-order weighted essentially
non-oscillatory (WENO) scheme is modified by considering computational cells 
being cut by the liquid-gas interface.
The interface is reconstructed from a signed distance
function $\phi$, i.e. the level set function, which gives the zero
level set for the location of interface \cite{OsherSethain1988}. 
The advection of the interface is achieved by solving the
equation
\begin{equation}
\phi_t + \mathbf{v}\cdot \nabla \phi =0 \label{level-set}
\end{equation}
with the 5th-order upwind WENO scheme \cite{FedkiwAslamXu1999}.
The fluxes across interface are obtained by impose jump conditions,
where the interface velocity along the normal direction $\mathbf{N}$ and pressure 
are determined by solving the liquid-gas Riemann problem within a narrow band of width $2b$ across the interface.
One state of the Riemann problem is obtained form the real fluid state, 
the other state is extrapolated from the other fluid by solving the extending equation \cite{FedkiwAslamXu1999}.
Note that, while a common interface velocity along the normal direction 
is obtained after solving the Riemann problem, 
the velocity jump along the tangential direction $\tau$ is preserved.

We consider a typical case which is first proposed by Chang and Liou \cite{chang2007robust},
in which a two-dimensional water drop in air 
is driven by a planar Mach 3 shock wave. 
With the dimensional references given by 1 atmosphere, 1 kg/m$^{3}$
and 1 cm, the drop has density of $1000$ and diameter of $0.35$, 
and the pre-shocked air has density $1.2$ and undisturbed pressure $1$.
The computation has been carried 
out on equidistant grid with with $\Delta = 2.5\times 10^{-3}$.
Since no explicit disturbances are imposed on the initial interface,
the numerical errors take this role.   

\begin{figure}[p]
\begin{center}
\includegraphics[width=0.5\textwidth]{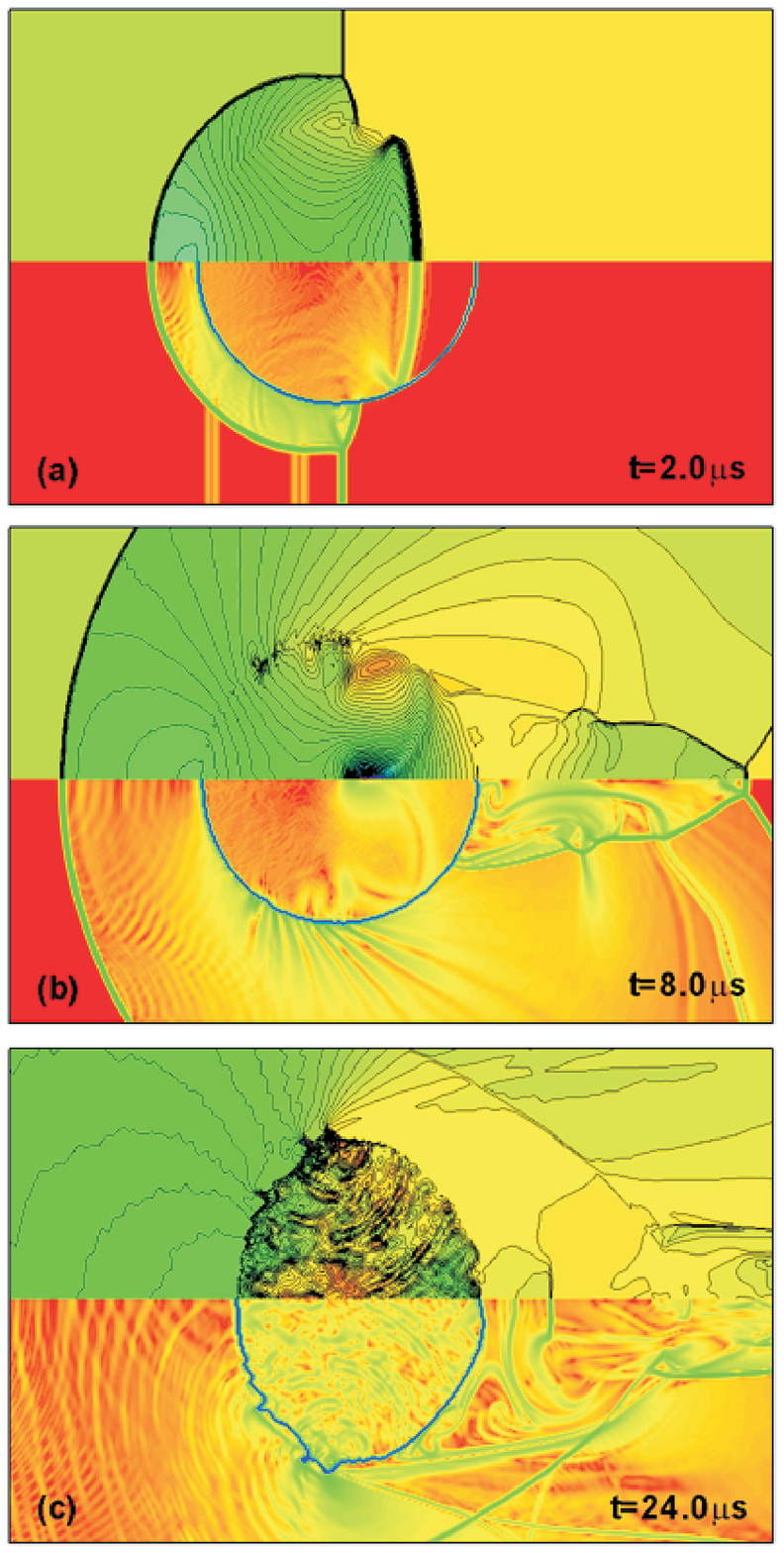}
\end{center}
\caption[]{Shock interacting with a two-dimensional water drop at $t = 0.0635$, 0.253 and 0.76: 
(upper) 55 contours of pressure, from -20 to 70;
(lower) 20 contours of flow speed, from 0.1 to 4; 
the thick white solid line indicates the air-water interface.} \label{gradient}
\end{figure}

Figure \ref{gradient} shows the contours of pressure and
flow speed at times $t = 0.0635$, 0.253 and 0.76 after the impact of the shock, 
representing the three stages of the development of the early-time interface instabilities.
At the first stage, as shown in Fig. \ref{gradient}a, the shock
diffracts around the drop just like around a
solid cylinder, the transmitted shock wave in the drop is weak and
travels faster than that in the air. 
While the pressure contours pass the air-water interface smoothly, 
a jump of pressure gradient has been built up at the upstream interface 
about the half way from the stagnation point to the equator with increasing magnitude.
The magnitude of pressure gradients from the air side and the water side 
is approximately the same.   
Meanwhile, a jump of tangential velocity has been built up at the upstream interface 
from a location nearer the stagnation point and achieves it maximum values at the equator.
At the second stage, as shown in Fig. \ref{gradient}b, 
while pressure contours pass
the air-water interface smoothly near the upstream front,
notable disturbances can be found increasing towards the equator.
It can be observed that the dominant disturbances starts at about 
the half way from the stagnation point to the equator, 
corresponding to the same location where the jump of pressure gradient starts, 
as shown in \ref{gradient}a.
Beyond the equator the pressure contours become smooth again.
While there is still a jump of pressure gradient,
the magnitude of pressure gradient in the air is much smaller than 
those at the upstream front and in the water drop.  
However, the magnitude of tangential velocity jump still increases
until it reaches the flow separation region.
At the third stage, as shown in Fig. \ref{gradient}c,
the drop is deformed considerably and the flow separation region extends to the equator. 
While wrinkles and discrete small droplets are produced at the locations 
where the pressure disturbances originate, the upstream front is relative smooth,
and has only small magnitude corrugations. 
The tangential velocity jump is smeared and turned into a shear layer with 
finite thickness above the wrinkles, 
which introduce tiny separation regions behind. 
In addition, it can be observed that small ripple-like acoustic waves bounce back and forth
between the primary bow shock and the drop,
and the acoustic waves within the drop produce a very complicated pattern.
Note that the simulations with much higher resolution than that used in this study
suggest that these above three stages and characteristics on pressure and velocity distribution 
are not grid-size dependent \cite{chang2013direct, han2014adaptive}. 
These predictions of interface deformation and fluctuation 
are in quite good agreement with the experimental results of the early-time of aerobreakup
\cite{theofanous2008physics, theofanous2011aerobreakup},
where the upstream front is relative smooth and 
the dominant disturbances starts at about the half way of the upstream interface toward the equator.

\begin{figure}[p]
\begin{center}
\includegraphics[width=0.6\textwidth]{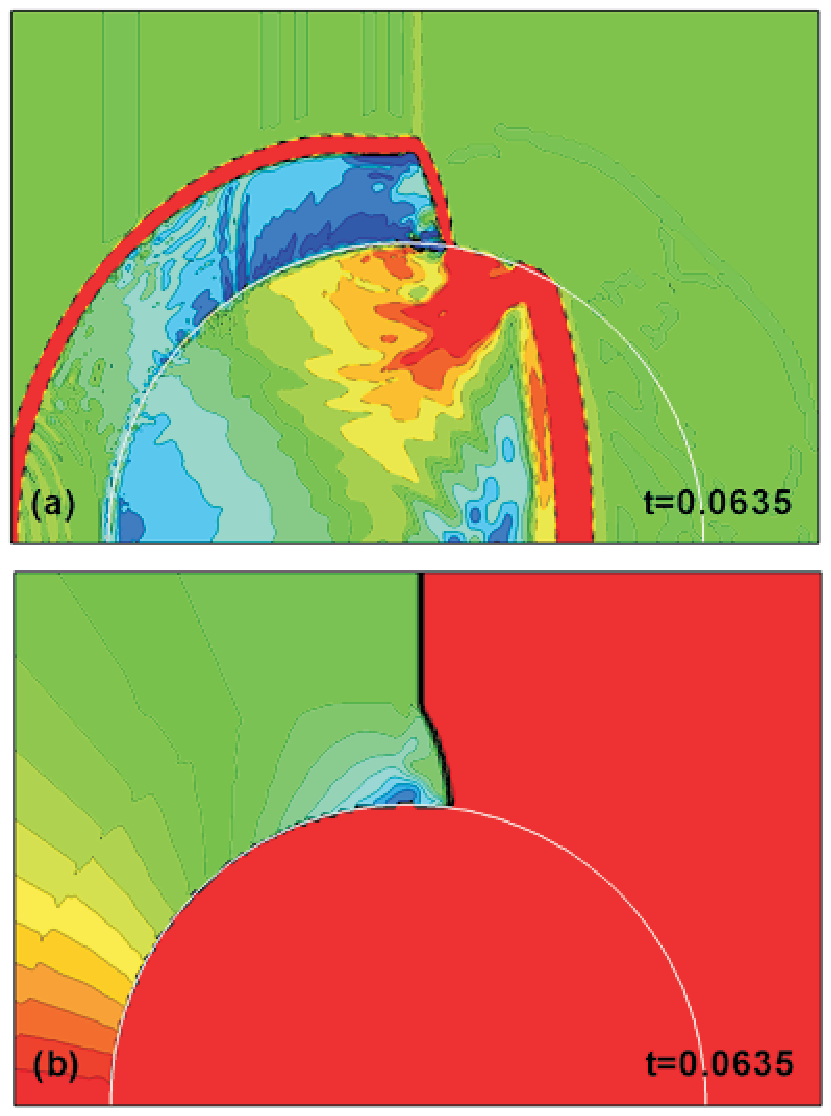}
\end{center}
\caption[]{Shock interacting with a two-dimensional water drop at $t = 0.0635$: 
(a) 11 contours (dash line for negative value) of pressure-gradient component normal to the interface 
$\rho a = \nabla p \cdot \mathbf{N}$, where $\mathbf{N}$ points from air to water, from -200 to 200;
(b) 20 contours of flow speed parallel to the interface $U = \mathbf{v} \cdot \mathbf{\tau}$, 
from 0.2 to 4; the thick white solid line indicates the air-water interface and 
the arrow in (a) points to the zero contour.} \label{gradient-1}
\end{figure}

The roles of RT and KH instabilities can be predicted by the linear-stability theory,
in which, when the density ratio $\rho_g/\rho_l << 1$, the growth rate is 
\begin{equation}
s = \frac{}{}\left[k^2 \frac{\rho_g U_{gl}}{\rho_g + \rho_l}+ 
k\frac{(\rho a)_{gl}}{\rho_g + \rho_l} \right]^{1/2}, \label{linear-theory}
\end{equation}
where the characteristic wavenumber of the initial disturbances $k$ can be assumed of order $\mathcal{O}(\Delta^{-1})$, 
$(\rho a)_{gl}$ are the jump of acceleration normal to the interface induced 
by the pressure gradient, 
$U_{gl} = U_g-U_l$ is the velocity jump tangential to the interface.
The first and second terms of Eq. (\ref{linear-theory}) are due to KH and RT instabilities, respectively. 
RT instability is dependent on the configuration of $(\rho a)_{}g$ and $(\rho a)_{l}$
It leads to instability only when $(\rho a)_{gl} > 0$, 
otherwise the interface is stabilized.
Note that the present analysis on RT instability is different from previous works 
\cite{harper1972breakup, joseph1999breakup},
in which an overall acceleration from air to water is assumed. 
To clarify the contributions of these two instabilities in the first stage,
the normal acceleration and tangential velocity are obtained and shown
in Fig. \ref{gradient-1}. 
It can be found that, near the stagnation point, 
one has $(\rho a)_{g}<0$, i.e. pressure increases from shock front to the stagnation point,
and $(\rho a)_{l}>0$, hence the interface is stabilized by this configuration of accelerations.
Since the tangential velocity around here is continuous 
and the effect of KH instability is also small,
such RT-based interface stabilization clearly explains
the smooth upstream front observed in experiments. 
It is also observed that, from the stagnation point towards the equator, 
both $(\rho a)_{g}$ and $(\rho a)_{l}$ change their signs.
On the upstream interface, one has the configuration of $(\rho a)_{g}>0$ and $(\rho a)_{l}<0$
beyond the zero contour of $(\rho a)_{l}$ (seen in Fig. \ref{gradient-1}a pointed by the arrow).
This suggests that the growth rate of RT instability increases substantially 
because the acceleration in both air and water destabilize the interface. 
On the other hand, the tangential velocity jump increases from 
the stagnation point toward the equator, 
and leads to increase of the growth rate of KH instability.
Note that, the development of the interface in the second stage
is in quite good agreement with these predictions.
As shown in Fig. \ref{gradient}b, notable disturbances start from 
the location where $(\rho a)_{l}$ changes its sign.
Note that this location of changing sign 
is at about the half way from the stagnation point to the equator, 
which is the same location where the jump of pressure gradient starts.
These observations clearly show that the essential contribution of RT instability.
However, since the growth rate of KH instability also increases along the upstream interface,
its contribution is not able to be ruled out.
For the present simulation, 
Eq. (\ref{linear-theory}) can be substituted with 
$k=\mathcal{O}(10^{-2})$, $\rho_g = 1.2$, typical $U_{gl}$ of 3 and 
$(\rho a)_{gl}$ from 200 to 300 along the upstream interface near equator.
It can be found that KH instability contributes 
more than RT instability for high-wavenumber disturbances. 
However, for disturbances with low- or intermediate-wavenumber,
RT instability can have comparable or higher growth rate.

\begin{figure}[ht!]
\begin{center}
\includegraphics[width=0.6\textwidth]{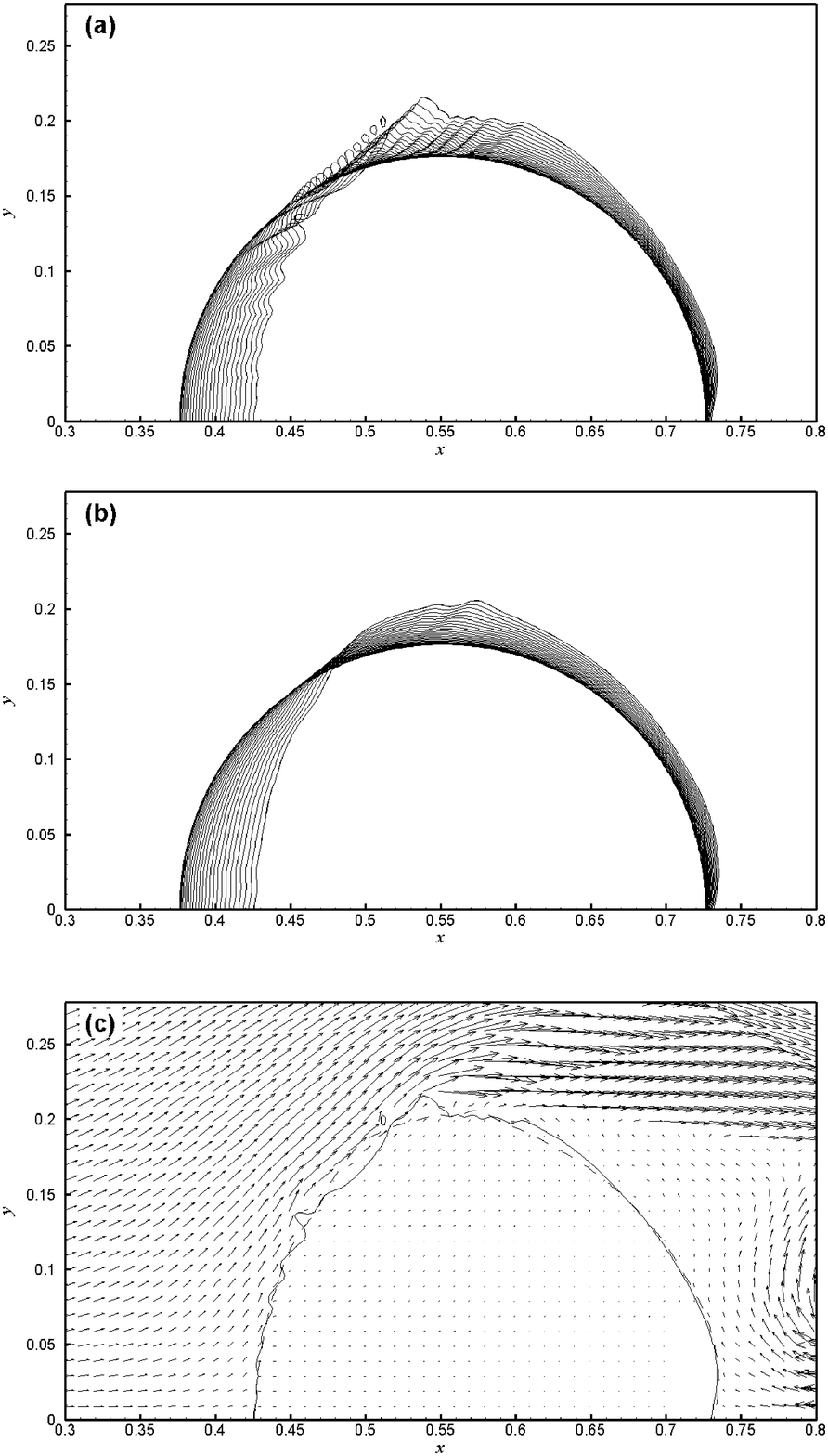}
\end{center}
\caption[]{The early-time (from $t=0$ to $0.76$) interface evolution: 
(a) the original simulation, (b) the modified simulation.
(c) The the velocity field of the original simulation 
and the interfaces of the original (solid line) 
and modified (dash line) simulations at $t=0.76$.}
\label{interface-evolution}
\end{figure}

In order to further verify the role of RT and KH instabilities,
a specially modified simulation is carried out.
In this simulation, 
the tangential component of the air velocity $\mathbf{v}_g$ near the interface is modified as 
\begin{equation}
\mathbf{v}_g \leftarrow (1-\alpha)\mathbf{v}_g +  \alpha\mathbf{v}_l + \alpha(\mathbf{v}_g - \mathbf{v}_l)\cdot\mathbf{N}, \label{modification}
\end{equation}
where $\alpha = (b-|\phi|)^2/b^2$ if $|\phi|<b$, otherwise $\alpha = 0$.
While keeping the other characteristics of the flow unchanged, 
this modification decreases the effect of KH instability considerably 
by eliminating the tangential velocity jump across the interface
and introducing a shear layer with thickness about $b$ above the interface.
Fig. \ref{interface-evolution} shows the evolution of the interface obtained in the original and modified simulations.
It can be observed that, compared to the original simulation, 
the evolution of the interface in the modified simulation is dominated with large scale deformation, and 
the high-wavenumber disturbances do not grow except for the regions near the equator, 
where the $(\rho a)_{gl}$ achieves its maximum, as shown in Fig. \ref{gradient-1}a.
These results clearly show that the development of the high-wavenumber disturbances 
is not sufficient if only with RT instability 
and KH instability must play an essential role. 
Note that both the original and modified simulations predict considerably radial motion 
at the upstream interface more than the downstream interface. 
Such behavior is in agreement with the experimental finding \cite{theofanous2011aerobreakup},
and suggests the such radial motion is mainly due to RT instability. 
A further comparison of the interfaces, as shown in Fig. \ref{interface-evolution}c, 
indicates that there is good agreement at the upstream and downstream fronts, 
which suggests that the development of RT instability 
is kept unchanged in the modified simulation for the regions 
where the growth rate of KH instability is small due to stagnation and flow-separation.

Note that, while the above analysis clarify that
the both RT and KH instabilities play considerable roles in the early-time of aero-breakup,
it is difficult to measure their contributions and their interplays 
quantitatively during the entire breakup process.
Though the linear-stability theory predicts that 
KH instability has larger growth rate for high-wave number disturbance,
the non-linear effect should be considered for the intermediate and late time.
As shown in Figs. \ref{gradient}c and \ref{interface-evolution}c, 
after the early-time instabilities,
the tangential velocity jump is smeared and turned into a shear layer. 
The thickness of the shear layer corresponds to the magnitude of interface fluctuations,
which leads to a dramatic decrease of the growth rate.
However, for non-linear RT instability, 
the multi-mode growth rate can be even lager than that in the linear regime 
\cite{sharp1984overview, kull1991theory, sohn2004vortex}.
In addition, non-linear RT instability leads to secondary KH instability, 
which may further destabilize the interface dramatically.

Another issue of the interface instabilities is the compressibility effects.
Previous studies suggest that the compressibility of fluid can either stabilize or destablilize the interface.
For KH instability, 
while the compressibility generally stabilize interface \cite{gerwin1968stability},
the sound-wave resonance effect in bounded domain can lead to 
increase of growth rate for high-wavenumber disturbances \cite{rajaee2007resonant}. 
The latter effect can explain the ripple-like waves between the primary bow shock 
and the upstream interface,
and the high-wavenumber pattern within the water drop.
The evidence that these waves are relevant to KH instability 
is that they are almost eliminated in the modified simulation (not shown here).
A simple prediction based on linear theory suggests that 
these waves do not contribute much due to the relative small growth rate  \cite{rajaee2007resonant}.
For RT instability, the compressibility influences the growth rate
by changing the density ratio \cite{bernstein1983effect, livescu2004compressibility}.
Since this change in aerobreakup is very small due to already very high density ratio,
it is very likely that the compressibility effect on RT instability is not obvious either.
However, to study the details of these effects, high-resolution simulations 
of intermediate- and late-time aerobreakup are required.
%
%Just because of unusual number of tables stacked at end
\bibliographystyle{plain}

\end{document}